\documentclass[aps,pra,10pt,twocolumn,showpacs,floatfix,superscriptaddress]{revtex4-1}
\pdfoutput=1

\usepackage{amssymb,amsmath}
\usepackage{array}
\usepackage{graphicx}

\def\d{\,\mathrm{d}}
\def\mb{\mu_{\mathrm{B}} }

\def\z{\textit{z}}
\def\Rb{${}^{87}\mathrm{Rb}$ }
\def\Na{${}^{23}\mathrm{Na}$ }
\def\adj{^\dagger}
\def\kink{e_{\vec{k}}}

\newcommand\deriv[2]{\frac{\partial #1}{\partial #2}}
\newcommand\tsex[2]{\frac{\partial #1}{\partial #2} #2}
\newcommand\mat[1]{\mathbf{#1}}

\begin{document}

\title{Three fluid hydrodynamics of spin-1 Bose-Einstein condensates}

\author{Gergely Szirmai}
\affiliation{Research Institute of Solid State Physics and Optics, Hungarian Academy  of Sciences, H-1525 Budapest P.O. Box 49, Hungary}
\author{P\'eter Sz\'epfalusy}
\affiliation{Research Institute of Solid State Physics and Optics, Hungarian Academy  of Sciences, H-1525 Budapest P.O. Box 49, Hungary}
\affiliation{Department of Physics of Complex Systems, E\"otv\"os University, H-1117 Budapest, P\'azm\'any P. s\'et\'any 1/A, Hungary}

\begin{abstract}
We study excitations of the spin-1 Bose gas at finite temperatures and in the presence of a not so strong magnetic field, or equivalently, when the gas sample is partially polarized. Motivated by the success of two-fluid hydrodynamics of scalar superfluids we develop a three-fluid hydrodynamic description to treat the low frequency and long wavelength excitations of the spin-1 Bose gas. We derive the coupled linear hydrodynamic equations of the three sounds and evaluate them numerically in a self-consistent mean field approximation valid for the dilute gas at the intermediate and critical temperature regions. In this latter region we identify the critical mode.
\end{abstract}


\maketitle

\section{Introduction}
\label{sec:intro}

Multicomponent dilute gases are fascinating systems that can exhibit ---at sufficiently low temperatures--- phase transitions where the interplay between a superfluid transition and the breaking of some other continuous symmetry can be observed. Maybe the simplest example is the Bose-Einstein condensation of the spin-1 Bose gas \cite{Ho2,OM}, where the system has a global SU(2) (spin rotation) symmetry as well as the usual U(1). For the weakly interacting gas Bose-Einstein condesation always infers the breaking of the SU(2) symmetry too, because the three component condensate wave function violates the spin rotation symmetry, even when spontaneous magnetization is zero. Depending on the sign and magnitude of the spin dependent part of the interaction a magnetic phase transition can occur prior to or at the onset of Bose condensation \cite{SzSz2,GK1,Kis-Szabo08a,Szirmai05a}. Note, that for relatively large interactions it has been shown recently that a pairing transition occurs instead of Bose condensation \cite{Natu11a}.

In the Bose condensed phase the global U(1) symmetry is always broken and correspondingly a Goldstone mode has to exist among the elementary excitations of the system. Though the spin rotation SU(2) symmetry is also broken by the condensate wave function two possiblities still remain. (1) A U(1) symmetry ---the rotation around the $z$-axis--- remains a symmetry, at least with the combination of the global U(1) gauge symmetry. Or, (2) even this kind of $z$-axis rotation symmetry is broken and the ground state has no remaining symmetry at all. In this latter case another Goldstone mode must exist. This kind of double symmetry breaking was illustrated in the Random Phase Approximation (RPA) for the polar spin-1 gas earlier, and the Goldstone modes have been identified amongst the quasiparticle excitations \cite{Kis-Szabo07}.

However, identifying the critical mode and studying its properties can be achieved in the hydrodynamic approximation which can be continued by scale invariance to the critical region. Such a hydrodynamic approximation, the celebrated 2-fluid hydridynamics of superfluids, is the basis of understanding the critical properties of liquid $^4\mathrm{He}$ even around the $\lambda$-transition and also serve for understanding the behavior of dilute, scalar Bose gases when interparticle collisions take place sufficiently frequently \cite{GNZ,Griffin97a,Zaremba98a}. In this paper we will follow the route originally developed for superfluid $^3\mathrm{He}$ \cite{Liu80a}, and derive a three-fluid model to describe the low energy excitations of the system. We will evaluate the thermodynamic quantitities, entering to the hydrodynamic equations, in a self-consistent mean-field approximation that treats the non condensed atoms as an ideal Bose gas above the spinor BEC.

The rest of the paper is organized as follows. In Sec. \ref{sec:ham}. the model is introduced with the help of the grand canonical Hamiltonian and its symmetry properties are discussed. In Sec. \ref{sec:phdiag}. the two possible phases and the phase diagram of the system is given in the magnetic field-temperature plane. Three fluid hydrodynamics is developed for the partially polarized phase in Sec. \ref{sec:3fluid}, with some details of the derivation moved to the Appendix. The thermodynamic quantities are evaluated in the Bogoliubov-Hartree approximation and are presented together with the results for the hydrodynamic excitations also in Sec. \ref{sec:3fluid}. Finally we summarize the results in Sec. \ref{sec:sum}.    

\section{Hamiltonian of the spin-1 Bose gas}
\label{sec:ham}

We consider a system of homogeneous, weakly interacting, dilute, spin-1 Bose gas
at ultralow temperatures and in a homogeneous magnetic field. The interparticle
interaction is modelled by s-wave scattering, i.e. we neglect the relatively weak
dipolar interaction of the gas. The grand-canonical Hamiltonian of the system
takes the following form:
\begin{multline}
\label{eq:ham}
  {\mathcal H}=\sum_{\genfrac{}{}{0pt}{2}{\vec{k}}{r,s}}
  \Big[(e_{\vec{k}}-\mu)\delta_{rs} -g \mb B\,
  (F_z)_{rs}\Big] a_r^\dagger(\vec{k})
  a_s(\vec{k})\\+\frac{1}{2V}\sum_{\genfrac{}{}{0pt}{2}{\vec{k}_1+
    \vec{k}_2=\vec{k}_3+\vec{k}_4}{r,s,r',s'}}a^\dagger_{r'}(\vec{k}_1)
  a^\dagger_r(\vec{k}_2)V^{r's'}_{rs}a_s(\vec{k}_3)a_{s'}
  (\vec{k}_4),
\end{multline}
where $a_r(\vec{k})$ and $a_r\adj(\vec{k})$ are the annihilation and creation
operators of plane wave states with momentum $\vec{k}$ and spin projection
$r$. The spin index $r$ refers to the eigenvalue of the \z-component of the
spin operator and can take values from $+,0,-$. Correspondingly $F_z=\mathrm{diag}(1,0,-1)$ is a 3x3 diagonal matrix. In Eq. \eqref{eq:ham} $\kink=\hslash^2k^2/(2M)$ refers to the kinetic energy of an atom, $\mu$ to the chemical potential, $g$ to the gyromagnetic ratio, $\mu_{\mathrm{B}}$ to the Bohr magneton, $B$ to the modulus of the homogeneous magnetic field. $V$ is the volume of the system and $V^{r's'}_{rs}$ the amplitude of the two particle interaction, given for spin-1 bosons by \cite{Ho2,OM,Stamper-Kurn2001a}:
\begin{equation}
  \label{eq:pseudopot}
    V^{r's'}_{rs}=c_n\delta_{rs}\delta_{r's'}+c_s(\vec{F})_{rs}
    (\vec{F})_{r's'},
\end{equation}
with $c_n=4\pi\hslash^2(a_0+2a_2)/(3M)$ , and $c_s=4\pi\hslash^2(a_2-a_0)/(3M)$.
The parameters $a_0$ and $a_2$ are the scattering lengths in the total hyperfine
spin channel zero and two, respectively. The constant $c_n>0$, while $c_s$ can
both be positive or negative, depending on the relative values of $a_0$ and $a_2$.
If $c_s<0$ the interaction tends to align the spins, while for $c_s>0$ a zero
net spin is energetically favorable (in the absence of a magnetic field). For
this reason systems with $c_s<0$ are referred as ferromagnetic systems, while
those with $c_s>0$ are called polar gases \cite{Ho2}. For example the ultracold
gas of \Rb atoms in the $f=1$ hyperfine state is ferromagnetic \cite{KBG}, while
the \Na gas (also in the $f=1$ hyperfine state) is polar \cite{Crea}.

The Hamiltonian \eqref{eq:ham} is invariant under the global $\mathrm{U}(1)$
symmetry transformation, which leads to particle number conservation. In the
case, when the external magnetic field is zero, the Hamiltonian is also invariant
under the global spin rotation $\mathrm{SU}(2)$, which leads to the conservation
of the total magnetization (spin) of the system. When a magnetic field is present
(breaking explicitly the rotational invariance), only another $\mathrm{U}(1)$
symmetry remains, namely the rotation around the \z-axis. These two
$\mathrm{U}(1)$ symmetries can be described either by $a_\pm\rightarrow a_\pm
e^{i\varphi_\pm}$ together with $a_0\rightarrow a_0
e^{i\frac{1}{2}(\varphi_++\varphi_-)}$, or equivalently by $a_r\rightarrow a_r
e^{i(\phi+r\theta)}$. It is easy to see that $\phi = \frac{1}{2}(\varphi_+ +
\varphi_-)$ corresponding to the symmetry generated by the particle number, and
$\theta = \frac{1}{2}(\varphi_+ - \varphi_-)$ corresponding to the symmetry
responsible for the conservation of the \z-component of the total spin. The
constraint due to the conservation of the total magnetization can be resolved
similarly as that of the conservation of particle number, i.e. with the introduction of a
Lagrange multiplier in the Hamiltonian \eqref{eq:ham}. This multiplier shows up
in the same way as the magnetic field does, therefore an effective magnetic field
can be introduced as a sum of the external magnetic field plus the Lagrange
multiplier. In the following $B$ will mean this kind of effective magnetic field even when the physical magnetic field is zero. In the following we assume that either such is the case, or the external magnetic field is so small that the quadratic Zeeman effect can be neglected.

\section{Phases of the spin-1 polar Bose gas}
\label{sec:phdiag}

\begin{figure}[!t]
\begin{center}
  \includegraphics{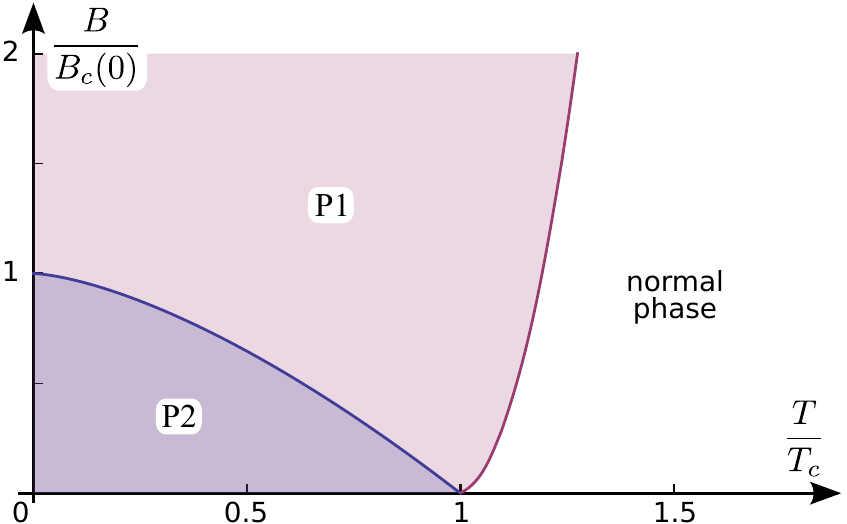}
  \caption{(Color online) The phase diagram of the polar spin-1 Bose gas in the Bogoliubov-Hartree approximation. $B_c(0)$ is the critical magnetic field at zero temperature, above which the condensate is fully polarized. Here we have chosen the coupling constants as $c_n=0.2\times M k_B T_c/\rho$ and $c_s=0.4\times c_n$.}
  \label{fig:phdiag}
\end{center}
\end{figure}
The weakly interacting spin-1 Bose gas can have two types of Bose-Einstein condensed phases. One is the fully polarized Bose-condensed phase that is realized by the thermodynamic ground state of a gas with $c_s<0$ (ferromagnetic coupling), or even when $c_s>0$ but the gas sample is originally polarized (or equivalently when the (effective) magnetic field is sufficiently large). 
This phase will be referred to as the P1 phase, in analogy to the similar phase of $^3\mathrm{He}$. In this phase the condensate wave function is single component, and correspondingly only one of the two global symmetries are broken, namely the symmetry parametrized by the $\varphi_+$ angle. In the case of a \emph{polar gas} and starting from the $P1$ phase, when the effective magnetic field is lowered, the system undergoes a ``second'' Bose--Einstein condensation. The condensate wave function is no longer single component and the other remaining global symmetry becomes broken too \cite{Kis-Szabo07}. This phase will be called as the P2 phase and one can assume that the condensate wave function has two nonzero components. We illustrate the phase diagram of polar spin-1 gas in Fig. \ref{fig:phdiag} in the self-consistent Bogoliubov-Hartree approximation, described later. Note, that the point $(T=T_c, B=0)$ is a special one, where two critical lines cross each other, and a renormalization group analaysis around $4-\epsilon$ dimensions to leading order in $\epsilon$ show that there is no stable fixed point \cite{Szirmai06a}.

\section{Three fluid hydrodynamics in the P2 phase}
\label{sec:3fluid}

\subsection{Equations of motion}
\label{sec:eqmo}

The system has two conserved global charges associated with the symmetry of the internal states, namely the total number of particles and the \z-component of the total magnetization, therefore two continuity equations arise
\begin{subequations}
\begin{align}
\partial_t \rho + \vec{\nabla} \vec{g}_\rho =0,\label{eq:cont1}\\
\partial_t \sigma + \vec{\nabla}\vec{g}_\sigma=0,\label{eq:cont2}
\end{align}
\end{subequations}
where $\rho$ is the mass denisty and $\sigma$ is the spin denisty (multiplied with the mass of the particles $M$). Correspondingly, $\vec{g}_\rho$ is the total current (momentum) density and $\vec{g}_\sigma$ is the spin current density.

In the P2 phase the Bose-Einstein condensates has two components, one is in the $+$ and the other is in the $-$ spin components. Directly following from the corresponding Gross-Pitaevskii equations and from Galilean invariance we have two independent superfluid components with their velocities obeying
\begin{subequations}
\begin{align}
M\partial_t\vec{v}_{s,+}+\vec{\nabla}(\mu+b)=0,\label{eq:vs+}\\
M\partial_t\vec{v}_{s,-}+\vec{\nabla}(\mu-b)=0.\label{eq:vs-}
\end{align}
\end{subequations}
The effect of fluctuations in the chemical potential $\mu$, and in the magnetic energy $b=g\mu_B B$ is nicely exhibited in an alternative representation. Instead of $\vec{v}_{s,+}$ and $\vec{v}_{s,-}$ one can introduce their linear combinations
\begin{subequations}
\label{eqs:newvels}
\begin{align}
\vec{v}_\rho&=\frac{\vec{v}_{s,+}+\vec{v}_{s,-}}{2},\\
\vec{v}_\sigma&=\frac{\vec{v}_{s,+}-\vec{v}_{s,-}}{2}.
\end{align}
\end{subequations}
Clearly $\vec{v}_\rho$ is the superfluid velocity for the density current and $\vec{v}_\sigma$ is the superfluid velocity for the spin current. They obey the following equations of motion
\begin{subequations}
\label{eqs:supveleqmo}
\begin{align} 
M\partial_t\vec{v}_\rho+\vec{\nabla}\mu=0,\\
M\partial_t\vec{v}_\sigma+\vec{\nabla}b=0.
\end{align}
\end{subequations}

Momentum conservation yields the well known Euler equation:
\begin{equation}
\label{eq:euler}
\partial_t\vec{g}_\rho+\vec{\nabla}p=0,
\end{equation} 
with $p$ being the pressure field. Another equation of motion is provided by the entropy transport of the normal component of the fluid:
\begin{equation} 
\label{eq:entropycont}
\partial_t s +\vec{\nabla}(s\vec{v}_n)=0,
\end{equation}
with $s$ being the entropy density of the gas.

To close the set of hydrodynamic equations one has to relate the complete density- and spin currents to the 3 components of the superfluid, i.e. to the normal component and to the two distinct superfluid ones. The only linear relation consistent both with the Maxwell relations detailed in the Appendix and with the zero temperature results one has to choose them to be
\begin{subequations}
\begin{align}
\vec{g}_\rho=\rho_n\vec{v}_n+\rho_s\vec{v}_\rho+\sigma_s\vec{v}_\sigma,\\
\vec{g}_\sigma=\sigma_n\vec{v}_n+\sigma_s\vec{v}_\rho+\rho_s\vec{v}_\sigma,
\end{align} 
\end{subequations}
with $\rho=\rho_n+\rho_s$ and $\sigma=\sigma_n+\sigma_s$. One can also split $\rho_s$ and $\sigma_s$ according to
\begin{subequations} 
\label{eqs:supsplit}
\begin{align} 
\rho_s&=\rho_{s,+}+\rho_{s,-},\label{eq:supdenssplit}\\
\sigma_s&=\rho_{s,+}-\rho_{s,-},\label{eq:supspindenssplit}
\end{align}
\end{subequations}
where $\rho_{s,+}$ and $\rho_{s,-}$ are the superfluid densities in spin components $+$, and $-$, respectively.

\subsection{The wave equations}

With the help of the equation of continuity for the density \eqref{eq:cont1} and with the Euler equation \eqref{eq:euler} one can derive the first of the three coupled wave equations:
\begin{equation}
\label{eq:pressurewave}
\partial_t^2\rho-\nabla^2 p =0.
\end{equation}
The second equation follows from the continuity equation for the spin density \eqref{eq:cont2} by combining it with the Euler equation \eqref{eq:euler} and the equations of motion for the superfluid velocities \eqref{eqs:supveleqmo} and the Gibbs-Duham relation. It reads as
\begin{multline}
\label{eq:spinwave}
\partial_t^2 \sigma-\frac{\sigma}{\rho} \nabla^2 p-\tilde{s}\frac{\sigma_n\rho_s-\rho_n\sigma_s}{\rho_n}\nabla^2 T\\
-\frac{\sigma(\sigma_n\rho_s-\rho_n\sigma_s)-\rho(\sigma_n\sigma_s-\rho_n\rho_s)}{\rho\rho_n M}\nabla^2 b=0,
\end{multline}
where we have introduced the $\tilde{s}=s/\rho$, the entropy per unit mass. The final equation is obtained from the entropy continuity equation \eqref{eq:entropycont} in combination with the Euler equation \eqref{eq:euler} and the equations of motion of the superfluid velocities \eqref{eqs:supveleqmo}:
\begin{equation} 
\label{eq:entropywave}
\partial_t^2\tilde{s}-\tilde{s}^2\frac{\rho_s}{\rho_n}\nabla^2 T-\tilde{s}\frac{\sigma\rho_s-\rho\sigma_s}{\rho\rho_nM}\nabla^2b=0.
\end{equation}

Close to equilibrium the fluctuation of the three densities can be expressed with the help of the fluctuations of the intensive parameters
\begin{subequations}
\label{eqs:lintrans}
\begin{align} 
\partial_t^2\rho&=\bigg(\frac{\partial\rho}{\partial p}\bigg)\partial_t^2p+\bigg(\frac{\partial\rho}{\partial T}\bigg)\partial_t^2T+\bigg(\frac{\partial\rho}{\partial b}\bigg)\partial_t^2b,\\
\partial_t^2\sigma&=\bigg(\frac{\partial\sigma}{\partial p}\bigg)\partial_t^2p+\bigg(\frac{\partial\sigma}{\partial T}\bigg)\partial_t^2T+\bigg(\frac{\partial\sigma}{\partial b}\bigg)\partial_t^2b,\\
\partial_t^2\tilde{s}&=\bigg(\frac{\partial\tilde{s}}{\partial p}\bigg)\partial_t^2p+\bigg(\frac{\partial\tilde{s}}{\partial T}\bigg)\partial_t^2T+\bigg(\frac{\partial\tilde{s}}{\partial b}\bigg)\partial_t^2b.
\end{align}
\end{subequations}
Combining Eqs. \eqref{eq:pressurewave}, \eqref{eq:spinwave}, \eqref{eq:entropywave}, and \eqref{eqs:lintrans} one gets a closed set of wave equations for the quantities $p$, $T$ and $b$ describing density waves, temperature waves and magnetization waves. When looking for the plane-wave solutions one uses
\begin{equation} 
\left(\begin{array}{c}p\\T\\b\end{array}\right)=\left(\begin{array}{c}p_0\\T_0\\b_0\end{array}\right)e^{i(\vec{k}\vec{r}-c\,k\,t)},
\end{equation} 
and the plane wave amplitudes and dispersions are obtained from the eigenvector problem:
\begin{equation} 
\label{eq:eigprob}
\mat{M}\left(\begin{array}{c}p_0\\T_0\\b_0\end{array}\right)=0,
\end{equation} 
where
\begin{widetext}
\begin{equation} 
\label{eq:hydmat}
\mat{M}=
\left(
\begin{array}{c c c}
\Big(\frac{\partial\rho}{\partial p}\Big)c^2-1 & \Big(\frac{\partial\rho}{\partial T}\Big)c^2 & \Big(\frac{\partial\rho}{\partial b}\Big) c^2\\
\Big(\frac{\partial\sigma}{\partial p}\Big)c^2-\frac{\sigma}{\rho} & \Big(\frac{\partial\sigma}{\partial T}\Big)c^2-\tilde{s}\frac{\sigma_n\rho_s-\rho_n\sigma_s}{\rho_n} & \Big(\frac{\partial\sigma}{\partial b}\Big)c^2-\frac{\sigma(\sigma_n\rho_s-\rho_n\sigma_s)-\rho(\sigma_n\sigma_s-\rho_n\rho_s)}{\rho\rho_nM}\\ 
\Big(\frac{\partial\tilde{s}}{\partial p}\Big)c^2&\Big(\frac{\partial\tilde{s}}{\partial T}\Big)c^2-\tilde{s}^2\frac{\rho_s}{\rho_n}&\Big(\frac{\partial\tilde{s}}{\partial b}\Big)c^2-\tilde{s}\frac{\sigma\rho_s-\rho\sigma_s}{\rho\rho_nM}
\end{array}
\right)
\end{equation}
\end{widetext}
The dispersion follows from $\det\mat{M}=0$ which is a cubic equation for $c^2$ and gives the velocities of the first sound, the quadrupolar mode and the second sound.

\begin{figure*}[tb!]
\begin{center}
  \includegraphics{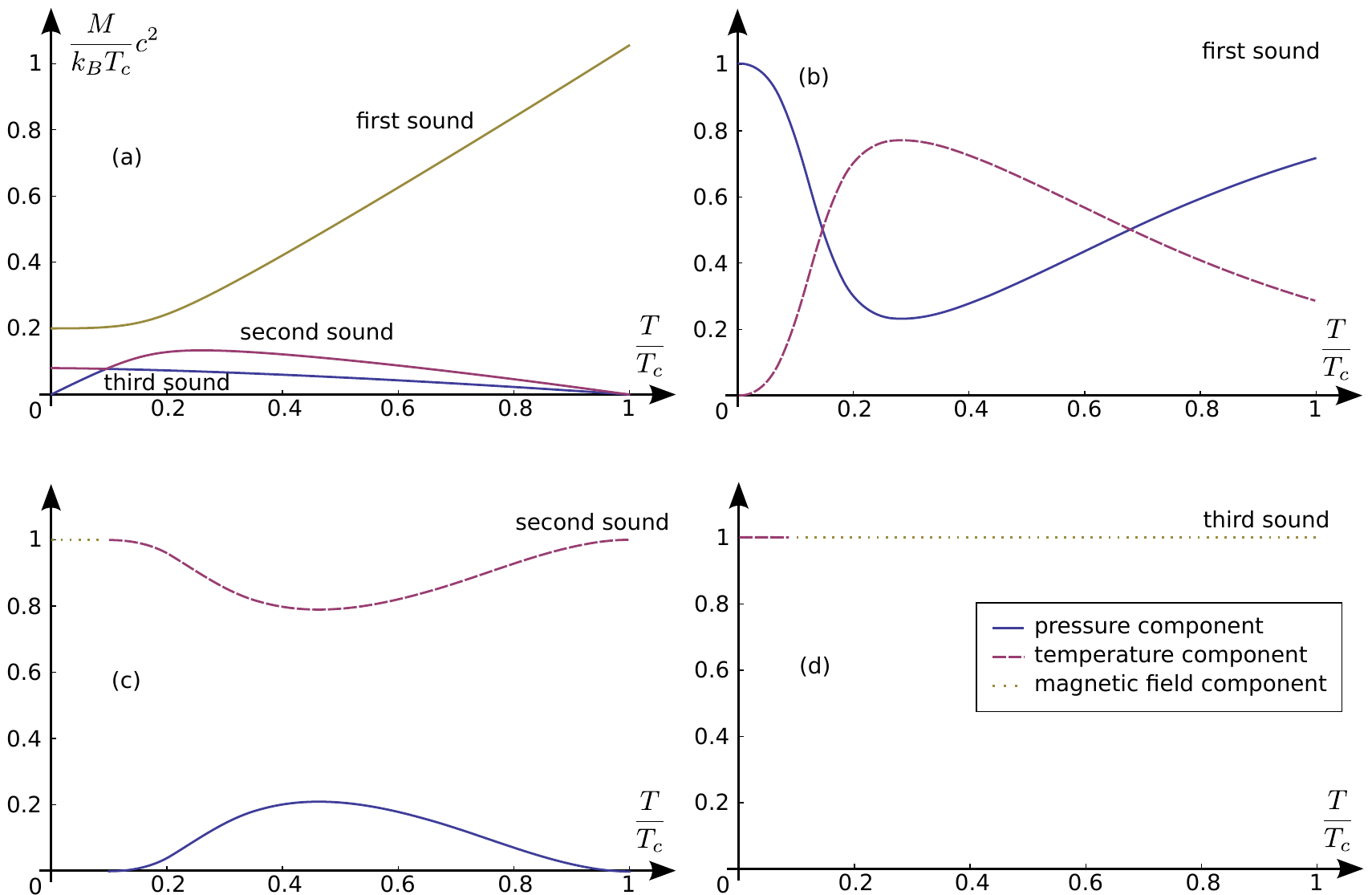}
  \caption{(Color online) The speeds of sound versus temperature at $B=0$ (a). The relative values of the three components of the eigenvectors corresponding to the given hydrodynamic mode: the first sound (a), the second sound (b) and the third sound (c). The coupling constants are chosen to be the same as in Fig. \ref{fig:phdiag}.}
  \label{fig:zerot}
\end{center}
\end{figure*}
\begin{figure*}[tb!]
\begin{center}
  \includegraphics{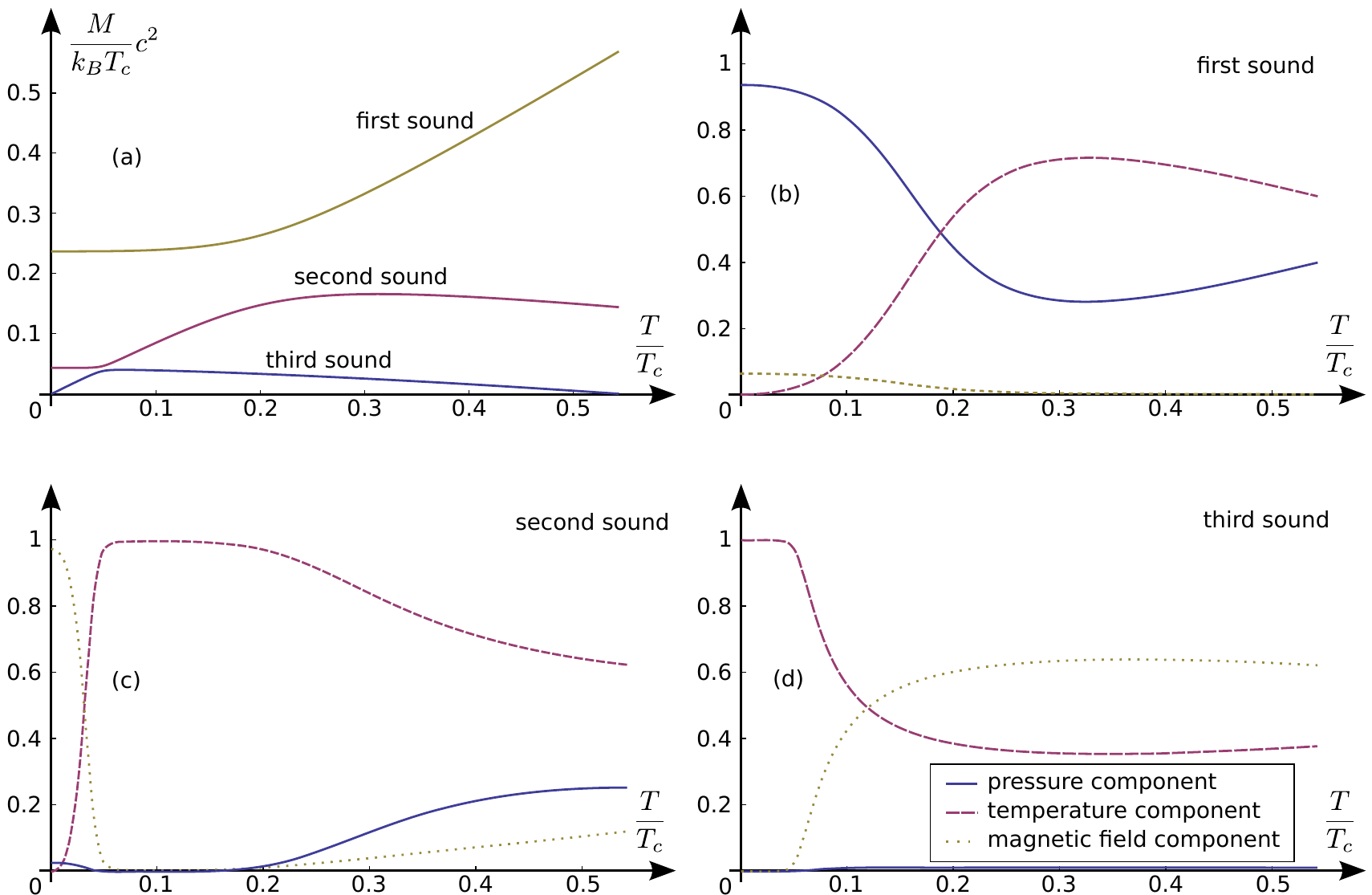}
  \caption{(Color online) The speeds of sound versus temperature at $B=0.6 B_c(0)$ (a). The relative values of the three components of the eigenvectors corresponding to the given hydrodynamic mode: the first sound (a), the second sound (b) and the third sound (c). The coupling constants are chosen to be the same as in Fig. \ref{fig:phdiag}.}
  \label{fig:fint}
\end{center}
\end{figure*}

\subsection{Evaluation of the thermodynamic quantities}
\label{ssec:eval}

In experiments made with $^{23}\mathrm{Na}$ atoms the typical ratio of the coupling constants is $\epsilon\equiv c_s/c_n\approx3\times10^{-2}\ll1$. In this limit the cubic equation for the speed of sound resulting from the vanishing of the determinant of Eq. \eqref{eq:hydmat} factorizes into a quadratic equation: the first and second sound modes, and to a separate linear equation with solution in the order of $\epsilon$: the quadrupolar spin-wave mode. This kind of separation is a generic feature of the dilute Bose gas and results from the behavior of the thermodynamic derivatives appearing in Eq. \eqref{eq:hydmat}. We are going to illustrate this feature in the Bogoliubov-Hartree approximation but keep the value of $\epsilon$ general in order to study the hybridization of the different sound waves.  

The advantage of the Bogoliubov-Hartree approximation is that it is an extension of the Bogoliubov approximation to fininte temperatures with accounting interaction effects with noncondensed atoms in a way which leads to a conserving and gapless approximation together with a continous phase transition  \cite{Kis-Szabo07}.

In this approximation the chemical potential and the magnetic field is expressed as
\begin{subequations} 
\begin{align} 
\mu=c_n \frac{\rho}{M},\\
b=c_s\frac{\sigma}{M}.
\end{align}
\end{subequations}
The normal density of spin component $r$ reads as
\begin{equation} 
\rho_{n,r}=\frac{M\zeta(3/2)}{\Lambda^3},
\end{equation}
where $\Lambda=\sqrt{2\pi\hslash^2/(Mk_BT)}$ is the de Broglie wavelength, and $\zeta(s)$ is the Riemann-zeta function. The total densities then
\begin{subequations} 
\begin{align}
\rho_n&=\rho_{n,+}+\rho_{n,0}+\rho_{n,-}=\frac{3M\zeta(3/2)}{\Lambda^3}=\rho\,t^{3/2},\\
\rho_s&=\rho_{s,+}+\rho_{s,-}=\rho\big(1-t^{3/2}\big),\\
\sigma_n&=\rho_{n,+}-\rho_{n,-}=0,\\
\sigma_s&=\rho_{s,+}-\rho_{s,-}=\sigma=\frac{M\,b}{c_s},
\end{align}
\end{subequations}
with $t=T/T_c$ is the reduced temperature. $T_c$ is the temperature of Bose-Einstein condensation of the spin-1 Bose gas in zero magnetic field (or with zero magnetization).
\begin{equation} 
T_c=\frac{2\pi\hslash^2}{k_B M}\bigg(\frac{\rho}{3\zeta(3/2) M}\bigg)^{2/3}
\end{equation}

The grand canonical thermodynamic potential in this approximation is
\begin{equation} 
\Phi(T,V,\mu)=-\frac{k_B T V}{\Lambda^3}3\zeta(5/2)-\frac{V}{2 M^2}\big(c_n \rho^2 + c_s \sigma^2\big).
\end{equation}
The pressure $p=-\Phi/V$, and its thermodynamic derivatives
\begin{subequations} 
\begin{align} 
\bigg(\frac{\partial p}{\partial T}\bigg)_{\rho,b}&=s=\frac{5}{2}\frac{k_B}{\Lambda^3}3\zeta(5/2),\\
\bigg(\frac{\partial p}{\partial \rho}\bigg)_{T,b}&=c_n\frac{\rho}{M},\\
\bigg(\frac{\partial p}{\partial b}\bigg)_{T,\rho}&=\frac{\sigma}{M},\\
\bigg(\frac{\partial \sigma}{\partial T}\bigg)_{\rho,b}&=0,\\
\bigg(\frac{\partial \sigma}{\partial \rho}\bigg)_{T,b}&=0,\\
\bigg(\frac{\partial \sigma}{\partial b}\bigg)_{T,\rho}&=\frac{M}{c_s}. \label{eq:susc}
\end{align}
\end{subequations}
The specific heat under constant volume is
\begin{equation} 
c_V=T\bigg(\frac{\partial s}{\partial T}\bigg)_{\rho,b}=\frac{3}{2}s.
\end{equation}

\subsection{Results and discussion}
\label{ssec:resdis}

The three solutions of $\det \mat{M}=0$ provides the three speeds of sound of the hydrodynamic modes, which we will call here as first, second and third sound. To identify which speed corresoponds to which physical excitation mode one has to look for the eigenvectors, i.e. the solution of Eq. \eqref{eq:eigprob}, corresponding to the specific speed of sound substituted into the matrix $\mat{M}$. In general the hydrodynamic modes hybridize, and their roles exchange. In Fig. \ref{fig:zerot} (a) we plot the three speeds of sound at the whole temperature range for zero magnetization (or equivalently at zero magnetic field). At this specific choice of zero magnetization the phase transition happens at $T_c$ and the P2 phase goes directly to the normal phase (compare with Fig. \ref{fig:phdiag} at $B=0$). This transition is characterized by the simoltaneous vanishing of the condensates both in $r=+$ and in $r=-$. Correspondingly two Goldstone modes are present, which are the second and third sounds. The speed of the first sound stays finite at this transition. On subfigures Fig. \ref{fig:zerot} (b)--(d) we demonstrate the amplitudes of the fluctuations of the sound modes. It can be seen that the first sound is basicly a pressure wave when we are far away from the point of the hybridization of the first and second sounds. The second sound is mainly temperature wave after the point of crossing with the third sound and not too close to the hybridization with the first sound. The third sound is a typical magnetic field wave after the crossing with the second sound. It is also very nicely demonstrated that the nature of the second and third sound changes at the crossing point.

On Fig. \ref{fig:fint} the same quantities are plotted but for finite magnetic field (or finite magnetization). It is visible from the figure, that it is the quadrupolar mode that goes soft at the P1-P2 transition point, thus it plays the role of the Goldstone mode. The second sound stays finite, since it only softens when the BEC gives way to the normal phase, that is only at the P1-normal transition point, which is not discussed in this paper. The first sound remains unmodified by the change of the magnetic field, since it is basically a pressure wave. The second and third sounds are hybridizing in this case and the crossing is resolved to an avoided one in the presence of the magnetic field, the explicit breaking of the rotational symmetry.

In the limit when $\epsilon=c_s/c_n\ll1$ the pressure and temperature waves separate from the quadrupolar spin wave. This limit can be treated by looking for solutions of $\det \mat{M}=0$ in the order of $\epsilon$. To leading order this gives the following solution:
\begin{subequations} 
\begin{equation} 
\label{eq:quadmode}
c^2=\frac{k_B T_c \epsilon \gamma_c}{M}\frac{1-\omega^2+t^3-2t^{3/2}}{1-t^{3/2}},
\end{equation}
which is the dispersion of the quadrupolar spin wave. For the other solutions one looks for $c\sim\mathcal{O}(1)$, and to leading order gets
\begin{equation} 
\label{eq:firstsecsounds}
c^4-\frac{k_B T_c}{M}\bigg(\gamma_c+\frac{2}{3\sqrt{t}}\frac{s M}{k_B\rho}\bigg)c^2+\frac{2}{3}\bigg(\frac{k_B T_c}{M}\bigg)^2\gamma_c\Big(\frac{1}{\sqrt{t}}-t\Big)=0,
\end{equation}
\end{subequations}
which is quadratic in $c^2$ and its solution gives the first and second sounds which formally agrees with that of the scalar Bose gas \cite{Griffin97a}. Note, that in the $t\rightarrow0$ limit we get back the solutions $c_1^2=k_B T_c \gamma_c/M$ and $c_2^2=0$. Thus at low temperature the lower of the two solutions crosses the quadrupolar mode of Eq. \eqref{eq:quadmode}, which stays finite at zero temperature. We have introduced the following dimensionless parameters $\gamma_c=c_n\rho/(M k_BT_c)$, $\omega=Mb/(c_s\rho)=\sigma/\rho$.

We have evaluated the thermodynamic derivatives in the Bogoliubov-Hartree approximation, which gives even quantitatively good interpolation between zero temperature and the critical point as far as equilibrium properties are concerned. As a matter of fact at intermediate temperatures also for the scalar gas we have evaluated the two fluid hydrodynamic equations in the Bogoliubov-Hartree aproximation and found the speeds of sounds very close to (few \%) the Popov approximation discussed in detail in Ref. \cite{Griffin97a}.
In the very low temperature region, however, where $k_B T<\mu$, the thermodynamical quantities are dominated by phonon excitations and show a qualtitively different behavior. The big difference is that the speed of the temperature wave does not go to zero but rather to $c_2^2=c_1^2/3$. Therefore neither of the three sounds vanishes at zero temperature.  In the limit of $\epsilon\ll1$ the quadrupolar mode is always situated below the temperature wave; thus the third sound is always a quadrupolar spin wave, while the second sound is a temperature wave.

One can assume that the susceptibility given in Eq. \eqref{eq:susc} remains singular when $c_s\rightarrow0$ beyond the Bogoliubov-Hartree approximation. Then the term $c^2 (\partial\sigma/\partial b)$ can remain finite if $c^2$ is of $\mathcal{O}(c_s)$, and in the matrix \eqref{eq:hydmat} all terms containing $c^2$ can be neglected in this limit except the one multiplied by the susceptibility (first term of $M_{2,3}$). As a result one arrives at the velocity of the quadrupolar spin mode in this limit as
\begin{equation}
\label{eq:quadwavegen}
c^2=4\frac{\rho_{s,-}\,\rho_{s,+}}{M\rho_s}\bigg(\frac{\partial\sigma}{\partial b}\bigg)^{-1},
\end{equation}
where Eqs. \eqref{eqs:supsplit} have been used. One can easily convince oneself that approximating the superfluid densities by the corresponding condensate densities \eqref{eq:quadwavegen} agrees with the expression \eqref{eq:quadmode}. However, Eq. \eqref{eq:quadwavegen} is valid to any orders in $c_n$. By the same token one can write the generalization of Eq. \eqref{eq:firstsecsounds} as the solution of the equation $\det\mat{M}_{2,3}=0$, where $\mat{M}_{2,3}$ denotes the submatirx obtained from \eqref{eq:hydmat} by removing the second row and the third column. 

Concerning the dynamical critical properties, the P2$\rightarrow$P1 transition on the $T,B>0$ plane belongs to the same universality class as the $\lambda$-transition of liquid Helium \cite{Ferrel67a,Hohenberg77a}. Consequently the quadrupolar spin wave suffers a damping $D\,k^2$, where $D\propto \Delta^{-1/3}$ when approaching the transition line between the P2 and P1 phases along the thermodynamic path on which $\Delta$ measures the distance from criticality.

\section{Summary}
\label{sec:sum}

In this paper we have developed a three fulid hydrodynamical approach for studying the low energy excitations of the spin-1 Bose gas in the partially polarized Bose-Einstein condensed phase (P2). Such a low energy and low momentum approximation of the excitation spectrum is considered to be valid, when interparticle scattering events happen sufficiently frequently in order to have large enough regions in the gas sample where thermodynamic equilibrium can be assumed. This kind of approximation is a generalization of the two fluid hydrodynamics, which was applied to scalar Bose-Einstein condensates successfuly \cite{Griffin97a,Arahata11a} and is in sharp contrast to quantum hydrodynamic approximations, already studied for spinor Bose-Einstein condensates too \cite{Rodrigues09a,Kudo10a}. These latter aproaches are nonlinear equations, equivalent to the Gross-Pitaevskii equation and are valid only at (or very close to) zero temperature and where quasiparticle interaction can be assumed to occur very rarely.

In the partially polarized Bose-Einstein condensed phase of polar spin-1 Bose gases (P2 phase) the fluid mass density can be decomposed to three components: a normal mass density and two superfluid densities. The relative motion of these three fluids determine the low energy excitations of the system. We have calculated the three sound velocities governing these fluctuations. We have found, that the speed of sound of the magnetization wave is always below that of the temperature and presussure waves for sufficiently weak spin-spin interaction ($c_s\ll c_n$). However, if $c_s$ is in the same order as $c_n$, successive hybrydizations occur, and the nature of the excitation branches exchange their character. One can hope that with the help of magnetic and optical Feshbach resonances such hybridizations can be achieved experimentally, and the three fluid nature of polar spin-1 Bose condensates and its implications on the phase transition can be examined.

\section{Acknowledgements}

This work was supported by the Hungarian National Research Fund (OTKA T077629). G.Sz. also acknowledges support from the Hungarian National Office for Research and Technology under the contract ERC\_HU\_09 OPTOMECH and the Hungarian Academy of Sciences  (Lend\"ulet Program, LP2011-016).

\appendix

\section{Thermodynamic quantities}
\label{sec:thermoquant}

The internal energy per unit volume of the superfluid spin-1 system in the P2 phase should look like
\begin{equation}
\label{eq:energ1}
\d e=M T\d s + \mu \d \rho + b \d \sigma + \vec{v}_n \d\vec{g}_\rho + \vec{j}_+\d\vec{v}_+ + \vec{j}_- \d\vec{v}_-.
\end{equation}
The quantities $\vec{v}_+=(\hslash/M)\nabla\varphi_+$,
$\vec{v}_-=(\hslash/M)\nabla\varphi_-$ are the superfluid velocities of the $+$
and $-$ spin components, respectively, while $\vec{j}_+$ and $\vec{j}_-$ are
their conjugate fields. In the P1 phase the last term is ommited from Eq.
\eqref{eq:energ1}.
In an alternative treatment $\phi$ and $\theta$ can be introduced instead of $\phi_+$ and $\phi_-$. It has an advantage of some physical insight. In this formulation the internal energy can be given by
\begin{equation}
\label{eq:energ2}
\d e=M T\d s + \mu \d \rho + b \d \sigma + \vec{v}_n \d\vec{g}_\rho + \vec{j}_\rho\d\vec{v}_\rho + \vec{j}_\sigma \d\vec{v}_\sigma.
\end{equation}

To obtain further thermodynamic equations one has to consider the transformation properties regarding an infinitesimal Galieleian transformation, $\d\vec{w}$. The noninvariant quantities transform as follows:
\begin{subequations}
\label{eqs:gali}
\begin{align}
\d e&=\vec{g}_\rho\d \vec{w},\\
\d \vec{g}_\rho&=\rho\d \vec{w},\\
\d \vec{v}_\pm&=\d\vec{w}.
\end{align}
\end{subequations}
With their help, the momentum density can be identified as:
\begin{equation}
\vec{g}_\rho=\rho\vec{v}_n+\vec{j}_++\vec{j}_-.
\end{equation}

In the alternative representation of the superfluid velocities, according to Eqs. \eqref{eqs:newvels} it is clear, that one of the superfluid velocities introduced in Eq. \eqref{eq:energ2} transform like a velocity: $\d\vec{v}_\rho=\d\vec{w}$, however the other one is Galilean invariant: $\d\vec{v}_\sigma=0$. Therefore the momentum density is as follows:
\begin{equation} 
\label{eq:momeq2}
\vec{g}_\rho=\rho\vec{v}_n+\vec{j}_\rho.
\end{equation}
Close to thermodynamic equilibrium the currents can be expanded to linear order in the velocities:
\begin{subequations}
\label{eqs:linexp2}
\begin{align}
\vec{g}_\rho&=\tsex{\vec{g}_\rho}{\vec{v}_n}+\tsex{\vec{g}_\rho}{\vec{v}_\rho}+\tsex{\vec{g}_\rho}{\vec{v}_\sigma},\\
\vec{j}_\rho&=\tsex{\vec{j}_\rho}{\vec{v}_n}+\tsex{\vec{j}_\rho}{\vec{v}_\rho}+\tsex{\vec{j}_\rho}{\vec{v}_\sigma},\\
\vec{j}_\sigma&=\tsex{\vec{j}_\sigma}{\vec{v}_n}+\tsex{\vec{j}_\sigma}{\vec{v}_\rho}+\tsex{\vec{j}_\sigma}{\vec{v}_\sigma}.
\end{align}
\end{subequations}
Equation \eqref{eq:energ2} yields the following Maxwell relations:
\begin{subequations}
\label{eqs:Maxwell2}
\begin{align} 
\deriv{\vec{g}_\rho}{\vec{v}_\rho}&=-\deriv{\vec{j}_\rho}{\vec{v}_n},\\
\deriv{\vec{g}_\rho}{\vec{v}_\sigma}&=-\deriv{\vec{j}_\sigma}{\vec{v}_n},\\
\deriv{\vec{j}_\rho}{\vec{v}_\sigma}&=\deriv{\vec{j}_\sigma}{\vec{v}_\rho}.
\end{align}
\end{subequations}
The nine coefficients of the linear expansion \eqref{eqs:linexp2} are not independent. With the help of the Maxwell realtions \eqref{eqs:Maxwell2} and Eq. \eqref{eq:momeq2}
\begin{subequations} 
\begin{align}
\vec{j}_\rho&=A(\vec{v}_\rho-\vec{v}_n)+B\vec{v}_\sigma,\\
\vec{j}_\sigma&=B(\vec{v}_\rho-\vec{v}_n)+C\vec{v}_\sigma
\end{align}
hold for linear order. Choosing A,B,C to be consistent also with the zero temperature linearized hydrodynamic equations gives
\begin{align}
A&=\deriv{\vec{j}_\rho}{\vec{v}_\rho}=-\deriv{\vec{j}_\rho}{\vec{v}_n}=\rho-\deriv{\vec{g}_\rho}{\vec{v}_n}=\deriv{\vec{g}_\rho}{\vec{v}_\rho}\equiv\rho_s,\\
B&=\deriv{\vec{j}_\rho}{\vec{v}_\sigma}=\deriv{\vec{j}_\sigma}{\vec{v}_\rho}=-\deriv{\vec{j}_\sigma}{\vec{v}_n}=\deriv{\vec{g}_\rho}{\vec{v}_\sigma}\equiv\sigma_s,\\
C&=\deriv{\vec{j}_\sigma}{\vec{v}_\sigma}=\rho_s.
\end{align}
\end{subequations}

\bibliography{bose}
\bibliographystyle{apsrev4-1}

\end{document}